\documentclass[aps,letterpaper,twocolumn,prl,showpacs,nofootinbib]{revtex4-2}

\usepackage{amssymb}
\usepackage{amsmath,bm}

\usepackage[pdftex]{graphicx,color}
\usepackage[english]{babel}
\usepackage{hyperref}
\usepackage{enumitem}
\usepackage{epstopdf}
\usepackage{tikz}

\begin{document}

\title{Ultrashort dissipative Raman solitons in Kerr resonators driven with phase-coherent optical pulses}

\author{Zongda Li$^{1,2}$}
\author{Yiqing Xu$^{1,2}$}
\author{Sophie Shamailov$^{1,2}$}
\author{Xiaoxiao Wen$^{3}$}
\author{Wenlong Wang$^{3}$}
\author{Xiaoming Wei$^{3}$}
\author{Zhongmin Yang$^{3}$}
\author{St\'ephane Coen$^{1,2}$}
\author{Stuart G. Murdoch$^{1,2}$}
\author{Miro Erkintalo$^{1,2,}$}
\email{m.erkintalo@auckland.ac.nz}
\affiliation{$^1$Department of Physics, University of Auckland, Auckland 1010, New Zealand}
\affiliation{$^2$The Dodd-Walls Centre for Photonic and Quantum Technologies, New Zealand}
\affiliation{$^3$School of Physics and Optoelectronics, South China University of Technology, Guangzhou 510640, China}

\begin{abstract}
External driving of passive, nonlinear optical resonators has emerged over the past decade as a novel route for the generation of ultrashort optical pulses and corresponding broadband frequency combs. Whilst the pulse formation dynamics in such systems differ dramatically from those manifesting themselves in conventional mode-locked lasers, the demarcation between the two traditionally distinct paradigms has recently begun to blur, with demonstrations of hybrid systems incorporating both external driving and active media shown to offer specific advantages. Here we explore a new pathway for ultrashort pulse generation at the interface of externally-driven passive resonators and lasers. By leveraging the nonlinear Raman gain inherent to fused silica, we achieve deterministic generation of low-noise dissipative solitons with durations well below 100 fs via phase-coherent pulsed driving of resonators made of standard, commercially-available optical fibre. We explore and explain the physics of the new dissipative Raman soliton states, identifying scaling laws that govern the pulses' characteristics and that allow output repetition rates to be scaled at will without influencing the soliton duration. The scheme explored in our work enables the shortest ever pulses generated in resonators (active or passive) made from a single commercially-available optical fibre, and it has the potential to be transferred into a chip-scale format by using existing dispersion-engineered silica microresonators.
\end{abstract}

\maketitle
\section{Introduction}

Coherent external driving of passive, Kerr nonlinear optical resonators with continuous wave (cw) or pulsed laser light allows for the generation of ultrashort pulses of light known as dissipative Kerr cavity solitons (CSs)~\cite{wabnitz_suppression_1993, leo_temporal_2010,jang_ultraweak_2013, herr_temporal_2014, obrzud_temporal_2017, kippenberg_dissipative_2018, lilienfein_temporal_2019}. Whilst such pulses were first observed in macroscopic ring resonators made of standard single-mode optical fibre~\cite{leo_temporal_2010, jang_ultraweak_2013}, their application potential was only unlocked by later realisations in monolithic Kerr microresonators~\cite{herr_temporal_2014, obrzud_temporal_2017, kippenberg_dissipative_2018, yi_soliton_2015, joshi_thermally_2016, brasch_photonic_2016}. In particular, dissipative Kerr solitons are the time-domain representations of the coherent and broadband microresonator frequency combs, whose burgeoning applications range from telecommunications~\cite{marin-palomo_microresonator-based_2017, corcoran_ultra-dense_2020} and artificial intelligence~\cite{xu_11_2021, feldmann_parallel_2021} to astronomy~\cite{suh_searching_2019, obrzud_microphotonic_2019} and distance measurements~\cite{suh_soliton_2018, riemensberger_massively_2020}.

Why are applications of CSs restricted to microresonators and not fibre resonators? In addition to the potential of on-chip integration~\cite{gaeta_photonic-chip-based_2019}, the answer lies in the fundamental scaling laws that govern the solitons' characteristics~\cite{coen_universal_2013}. Specifically, fibre resonators typically have long lengths and modest nonlinearity and finesse, resulting in solitons with picosecond-scale durations. While durations below 200~fs can be achieved using judicious fibre-based dispersion management~\cite{nielsen_invited_2018,spiess_chirped_2021,dong_120-fs_2022}, these numbers are still substantially larger than what can be realised in state-of-the-art integrated microresonator systems. Conversely, the microresonator platforms that permit dissipative solitons with durations measured in tens of femtoseconds typically have very large (hundreds of gigahertz) free-spectral range~\cite{brasch_photonic_2016,moille_ultra-broadband_2021}, which makes them unsuitable for applications that require frequency combs with fine spacing (e.g. high-resolution spectroscopy). There is therefore a need to break the scaling laws that govern dissipative solitons in driven Kerr resonators in order to concomitantly realise ultrashort pulse durations and repetition rates in the sub-GHz to GHz range.

CSs are fundamentally different from pulses generated in mode-locked lasers~\cite{leo_temporal_2010, firth_buffering_2010}: they exist in \emph{passive} resonators, gaining their energy via instantaneous, phase-sensitive Kerr nonlinear interactions with the quasi-cw background on top of which they sit. Mode-locked lasers, in contrast, operate via stimulated emission provided by an active medium, with typical gain relaxation timescales orders of magnitude larger than other characteristic timescales of the system. Recently, however, the boundary between the active and passive paradigms of ultrashort pulse (and broadband comb) generation has begun to blur, with studies considering pulse generation in hybrid systems that incorporate both external coherent driving as well as material gain~\cite{bao_laser_2019, englebert_temporal_2021, columbo_unifying_2021,  rowley_self-emergence_2022}. But, of course, even resonators that are considered purely \emph{passive} can be associated with nonlinear optical processes that are akin to phase-insensitive stimulated emission -- such as stimulated Brillouin and Raman scattering. These processes have been leveraged in a number of studies to achieve cw lasing in microresonator systems~\cite{spillane_ultralow-threshold_2002, lee_chemically_2012}. They are also known to influence the dynamics and characteristics of CSs~\cite{milian_solitons_2015, webb_experimental_2016, wang_stimulated_2018}, as well as give rise to new forms of dissipative solitons~\cite{yang_stokes_2017, babin_multicolour_2014} and soliton interactions~\cite{volkel_intracavity_2022}.

In this Article, we experimentally and theoretically explore a new paradigm of ultrashort pulse generation at the interface of externally-driven passive resonators and lasers. Building upon a recent serendipitous discovery~\cite{xu_frequency_2021}, we show that the broadband stimulated Raman scattering (SRS) gain inherent to fused silica, in conjunction with phase-coherent pulsed pumping, allows for a new form of ultrashort dissipative Raman soliton, with record durations well below 100~fs achieved in fibre resonators made from standard telecommunications fibre and components. In stark contrast to early implementations of synchronously-pumped fibre Raman oscillators~\cite{kafka_fiber_1987, Gouveia-Neto_soliton_1988, keller_noise_1989, harvey_argon_1990}, we leverage a carefully chosen resonator dispersion and resonant, phase-coherent pumping to obtain phase-locked operation that completely overcomes the low coherence and comparatively broad pulses associated with earlier systems. We show that the resultant ultrashort Raman solitons can be excited deterministically with perfect fidelity, and that they can be sustained in resonators of varying lengths, thus offering a route to broadband frequency combs with bespoke line spacing. Our current experiments focus on fibre resonators and the generation of sub-GHz combs that could find immediate use in applications that benefit from fine comb spacing (e.g. dual-comb spectroscopy), but we envisage further that the Raman soliton states explored in our work will enable substantially improved  performance of GHz combs in integrated silica microresonators~\cite{lee_chemically_2012, yang_broadband_2016, fujii_dispersion_2020}.

\section{Results}

\begin{figure*}[!t]
	\centering
	\includegraphics[width = \textwidth, clip=true]{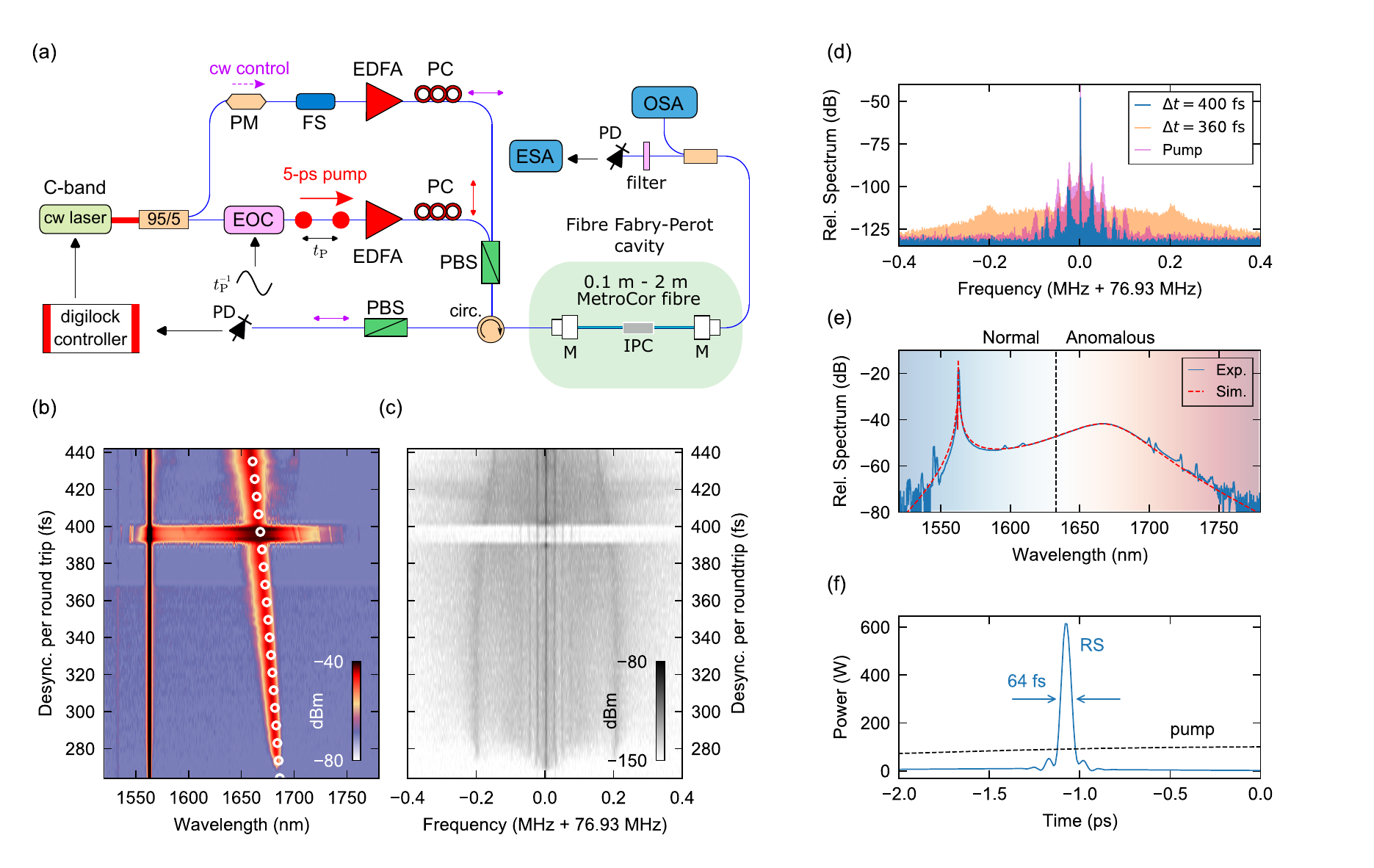}
	\caption{\textbf{Experimental demonstration of dissipative Raman soliton generation.} (a) Experimental setup. cw, continuous wave; PM, phase modulator; FS, frequency shifter; EDFA, erbium-doped fibre amplifier; PC, polarization controller; EOC, electro-optic comb generator; PBS, polarizing beam splitter; circ., circulator; M, mirror; IPC, in-line polarization controller; PD, photodetector; OSA, optical spectrum analyzer; ESA, electronic spectrum analyzer. (b) Optical spectra measured at the cavity output as a function of the desynchronization per round trip. Open white circles indicate the calculated wavelength that is group-velocity-matched to the input pulse train. (c) RF spectra corresponding to (b). (e) Example RF spectra measured with the desynchronization in the Raman soliton regime (blue), outside of that regime (orange), and the input pump pulse train (pink). (e) Example optical spectrum measured (blue solid curve) for a desynchronization $\Delta t \approx 400~\mathrm{fs}$; dashed red curve shows corresponding simulation result. (f) Blue solid curve shows the simulated temporal intensity profile of the spectrum shown in (b), whilst the black dashed curve shows the input pump profile for reference. All results obtained in a resonator with round trip length and finesse approximately 2.65 m and 150, respectively. See Methods for other parameters.}
	\label{fig1}
\end{figure*}

We begin by describing our experimental platform [Fig.~\ref{fig1}(a)] and by presenting illustrative results of Raman soliton formation. Our setup is built around a passive fibre Fabry-P\'erot resonator with zero-dispersion wavelength at 1633~nm that is formed from two dielectrically-coated mirrors that surround a single segment of non-zero dispersion-shifted optical fibre (Corning MetroCor) [see Methods]. We chose a Fabry-P\'erot (rather than a ring) architecture to allow more easily control the resonator length, yet note that the Raman solitons also manifest themselves in ring-type resonators~\cite{xu_frequency_2021}. The mirrors that form our resonator impose an upper limit of about 500 for the system's finesse, but our setup also gives us the option to introduce additional micro-bend losses when required [see Methods], thus permitting us to systematically study how the Raman soliton characteristics depend upon the resonator finesse. We emphasise that our resonator does not include any active gain medium.

The resonator is driven with a train of roughly 5~ps duration pulses generated from a narrow-linewidth cw laser via electro-optic comb (EOC) generation referenced to an RF signal generator; the frequency of the RF generator sets the repetition rate of the pulse train [see Methods]. Before the pulses are launched into the resonator, they are amplified with a C-band erbium-doped optical amplifier and their polarization is aligned along one of the eigenmodes of the resonator. In the first experiments to be discussed, the central wavelength of the cw laser is set to about 1562.5~nm which is in the region of normal group-velocity dispersion of the resonator. This wavelength is actively stabilized to coincide with a cavity resonance (zero detuning) by applying the Pound-Drever-Hall technique on a low-power cw probe beam that is orthogonally polarized with respect to the picosecond driving pulses~\cite{nielsen_invited_2018, li_experimental_2020}. This ensures that our experiment operates in the coherent regime with constant relative phase between the picosecond driving pulses and the field circulating inside the resonator.

The desynchronization $\Delta t = t_\mathrm{R}(\omega_\mathrm{P})-t_\mathrm{P}$ between the period of the driving pulse train ($t_\mathrm{P}$) and the resonator round-trip time at the driving frequency [$t_\mathrm{R}(\omega_\mathrm{P})$] is a key control parameter for the dynamics of interest. Figure~\ref{fig1}(b) shows experimental measurements of the optical spectrum at the output of a 2.65-m-long resonator as a function of the desynchronization, achieved by changing the RF clock signal that drives the EOC generator. As can be seen, when the magnitude of the desynchronization is sufficiently large, new spectral components that are red-shifted with respect to the injected pulse train are generated inside the resonator. These components originate from stimulated Raman scattering, as evidenced by their frequency down-shift of about 10~THz compared to the pump~\cite{agrawal_nonlinear_2019}. The exact frequencies $\omega$ of the components' spectral maxima are determined by a group-velocity-matching condition: the components that are most efficiently amplified are those with a group-velocity that yields a round-trip time equal to the period of the input pulse train, i.e., $t_\mathrm{R}(\omega)=t_\mathrm{P}$ [see circles in Fig.~\ref{fig1}(b) and Methods]. In other words, whilst the injected pulse train is desynchronous with respect to the pump frequency, the Raman components that emerge are those for which the driving is synchronous. We note that such group-velocity-matching is reminiscent of Raman-induced energy transfer between higher-order modes in resonators~\cite{yang_stokes_2017} and single-pass fibre systems~\cite{rishoj_soliton_2019,antikainen_fate_2019}, but occurs here by virtue of the desynchronization at the pump frequency.

A key feature of our scheme is that, whilst the injected pump field lies in the \emph{normal} dispersion regime, the group-velocity-matched Raman components reside in the \emph{anomalous} dispersion regime, thus permitting soliton formation. Indeed, the measurements depicted in Fig.~\ref{fig1}(b) reveal that, for desynchronizations within a small interval, considerable spectral broadening takes place. Concomitantly with the broadening of the optical spectrum, the RF spectrum collapses to a narrow structure around the fundamental beat tone [Fig.~\ref{fig1}(c)], signalling the emergence of a low-noise soliton state.  Figure~\ref{fig1}(d) and (e) show the RF and optical spectra measured for such a state, respectively. The RF spectrum replicates the corresponding spectrum of the injected pulse with little additional noise [see Supplementary Fig.~2 for the phase noise spectra], whilst the optical spectrum shows a broad spectral envelope that bridges the group-velocity-matched component at 1665~nm and the injected pump component at 1562.5~nm. Also shown in Fig.~\ref{fig1}(e) is the optical spectrum obtained from numerical simulations of the generalized Lugiato-Lefever equation~\cite{wang_stimulated_2018} using parameters corresponding to the experiments [see Methods]. The simulation results are in excellent agreement with the experimentally measured spectrum. The simulations additionally reveal that, in the time-domain, the intracavity field corresponds to a 64-fs-long pulse that sits atop a low-level pedestal [Fig.~\ref{fig1}(f)]. We must emphasise that the results in Fig.~\ref{fig1} were obtained with the pump peak power set to about 100~W, which implies that the average pump power required to support the Raman solitons in the experiments is as low as 40~mW. We also highlight that the Raman solitons are generated deterministically and with perfect fidelity: every time the pump desynchronization is tuned into the appropriate regime, a soliton is formed [see Supplementary Fig.~1].

\begin{figure}[!t]
	\centering
	\includegraphics[width = \columnwidth, clip=true]{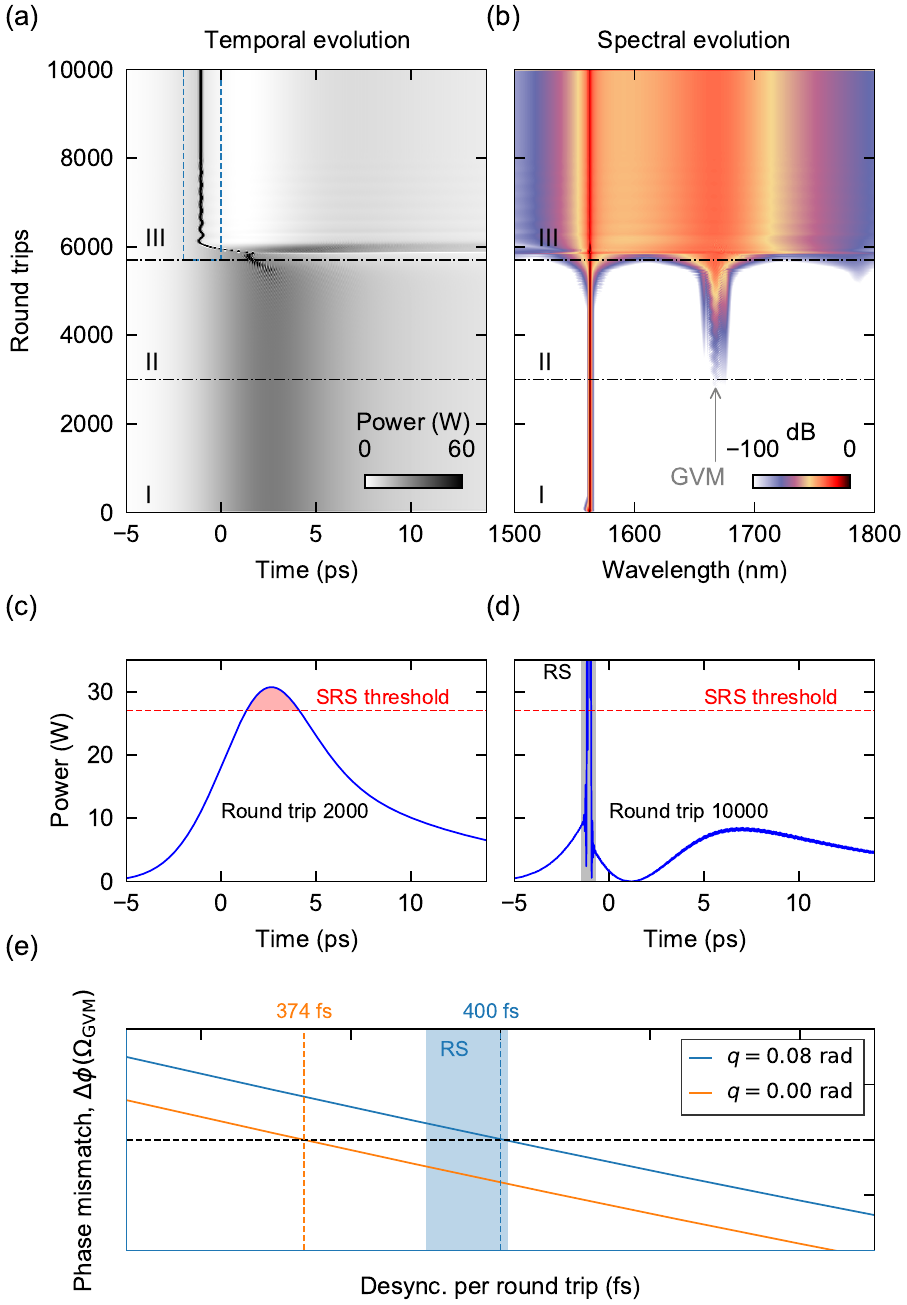}
	\caption{\textbf{Numerically simulated dynamics of Raman soliton formation.} (a) and (b) respectively show the round trip to round trip evolution of the intracavity temporal and spectral intensity profile, demonstrating the formation of a dissipative Raman soliton. The parameters are the same as those used in Fig.~\ref{fig1}(d) and (f) [see Methods]. The temporal dynamics are shown in the reference frame of the group-velocity-matched component. The dashed blue rectangle in (a) highlights the Raman soliton. GVM in (b) highlights the group-velocity-matched component as predicted from Eq.~\eqref{GVM}. (c, d) Intracavity profile (c) before and (d) after Raman soliton formation. The dashed horizontal lines show the SRS threshold [see Methods]. The shaded rectangle in (d) highlights the Raman soliton (RS). (e) Phase-mismatch [$\Delta\phi$, left-hand side of Eq.~\eqref{PM}] as a function of desynchronization $\Delta t$ with (blue curve) and without (orange curve) taking into account the nonlinear phase shift of a propagating soliton with peak power $P_\mathrm{s} \approx 600~\mathrm{W}$, i.e., $q = \gamma P_\mathrm{s}/2=0.08~\mathrm{rad\, m^{-1}}$, where $\gamma = 2.6~\mathrm{W^{-1}km^{-1}}$ is the Kerr nonlinearity coefficient. The shaded blue region indicates the Raman soliton existence range observed in experiments [see Fig.~\ref{fig1}(b)]. }
	\label{fig2}
\end{figure}

To gain more insights, Figs.~\ref{fig2}(a) and (b) respectively show the temporal and spectral growth dynamics of the dissipative Raman soliton of Fig.~\ref{fig1}(d) and (f) as obtained from numerical simulations starting from an empty cavity [see also Supplementary Video]. The dynamics can be divided into three regimes. In the first regime, a temporally broad and spectrally narrow pulse forms inside the resonator at the wavelength of the injected pulse train. In the absence of SRS, this state would be the steady-state of the system. However, because of SRS, the intracavity field at the injected wavelength amplifies a spectral component at the group-velocity-matched frequency (regime II). In the time-domain, the SRS gain is above threshold only over a small temporal interval around the peak of the intracavity field at the pump wavelength [see Fig.~\ref{fig2}(c) and Methods], due to which the amplification results in the break-up of the intracavity field about its temporal peak. From this break-up, an ultrashort soliton is seen to emanate (regime III), reaching a steady-state where the soliton sits at the leading edge of the intracavity pump pulse. Because of the pump-desynchronization (and the finite response time of SRS), the soliton is able to deplete the intracavity pump pulse in a way that pushes the Raman gain completely below threshold, thus enabling low-noise, steady-state operation whilst preventing the formation of a second soliton [Fig.~\ref{fig2}(d)].

\begin{figure*}[!t]
	\centering
	\includegraphics[width = \textwidth, clip=true]{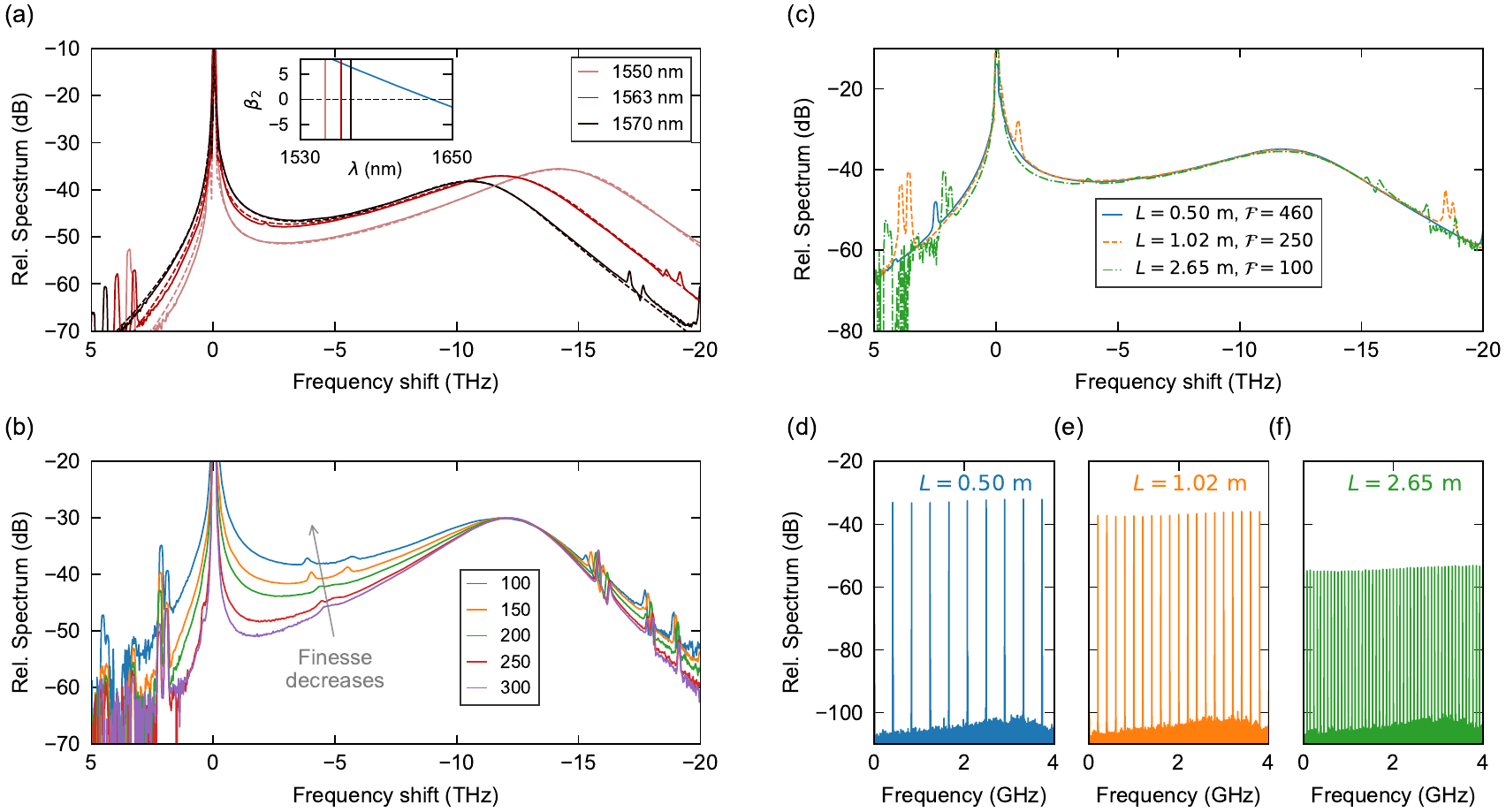}
	\caption{\textbf{Scaling laws of dissipative Raman solitons.} (a) Solid and dashed curves respectively show experimentally measured and numerically simulated optical spectra versus frequency shift from the pump for three different pump wavelengths as indicated. The resonator used had a round trip length of about 1~m and finesse of 400. Insets shows the group-velocity dispersion profile of the resonator. Pumping further away from the zero-dispersion wavelength yields an overall broader spectrum. (b) Raman soliton spectra measured for constant round trip length (2.65 m) but for various values of the cavity finesse (indicated in the legend). The spectral width increases with decreasing finesse. (c) Raman soliton spectra measured in three different resonators with approximately constant product between resonator finesse and length. (d)--(f) RF spectra corresponding to Raman soliton pulse trains generated in the different resonators used in (c). The results in (c)--(f) highlight that similar Raman solitons can be generated in various resonator configurations, thus allowing to realise ultrashort pulse trains with bespoke repetition.}
	\label{fig3}
\end{figure*}

Why does the low-noise dissipative Raman soliton only manifest itself for some specific pump desynchronizations? Extensive simulations reveal that, in addition to group-velocity-matching, the Raman component must be phase-matched to the intracavity field at the pump frequency. The two conditions can be written as [see Methods]
\begin{align}
\frac{\Delta t}{L}\Omega_\mathrm{R} + \hat{D}(\Omega_\mathrm{R}) + q = 0 & & \textnormal{Phase-matching,} \label{PM} \\
\frac{\Delta t}{L} + \hat{D}_1(\Omega_\mathrm{R}) = 0 & & \textnormal{GV-matching.} \label{GVM}
\end{align}
Here $L$ is the resonator round trip length, $\Omega_\mathrm{R} = \omega_\mathrm{R} - \omega_\mathrm{P}$ is the angular frequency shift between the Raman component and the pump, $q$ is a nonlinear phase shift, and $\hat{D}(\Omega)$ is the reduced dispersion of the fibre that forms the resonator:
\begin{equation}
\hat{D}(\Omega) = \sum_{k\geq 2} \frac{\beta_k}{k!}\Omega^k,
\end{equation}
with $\beta_k$ the Taylor series expansion coefficients of the fibre propagation constant $\beta(\omega)$ evaluated at the pump frequency $\omega_\mathrm{P}$. Finally, $\hat{D}_1(\Omega) = d\hat{D}/d\Omega$ is the group-velocity-mismatch relative to the pump frequency. To provide a physical basis for the significance of phase-matching, we note that the condition given by Eq.~\eqref{PM} essentially implies that the pump wavelength is equal to the wavelength of resonant dispersive waves emitted by the Raman soliton~\cite{skryabin_colloquium_2010}. This observation suggests that the emergence of the low-noise Raman states can be interpreted as a form of injection-locking between the solitons' dispersive wave and the residual pump background.

For typical resonator dispersion, and for a given nonlinear phase shift $q$, Eqs.~\eqref{PM} and Eqs.~\eqref{GVM} can be simultaneously satisfied only for a single desynchronization $\Delta t$. As an example, in Fig.~\ref{fig2}(e) we plot the phase-mismatch [left-hand side of Eq.~\eqref{PM}] as a function of desynchronization $\Delta t$ for resonator parameters corresponding to our experiments with the nonlinear phase shift set to zero (orange curve) and to that of the nonlinear phase shift of a propagating soliton with peak power $P_\mathrm{s}=600~\mathrm{W}$ [see Fig.~\ref{fig1}(f)]. As can be seen, the phase-mismatch crosses zero at $\Delta t \approx 400~\mathrm{fs}$, which is precisely the value corresponding to the experiments and simulations shown in Fig.~\ref{fig1}(d) and (f). In practice, the steady-state Raman soliton states can be sustained over a small range of desynchronizations (rather than a single value) due to the solitons' ability to compensate for small phase-mismatches by adjusting their peak power (and hence the nonlinear phase shift $q$).

Equations~\eqref{PM} and~\eqref{GVM} can be leveraged to qualitatively understand how the Raman soliton characteristics depend upon the resonator dispersion. Specifically, ignoring the nonlinear phase shift ($q = 0$), the conditions allow to predict the soliton's spectral shift from the input pump ($\Omega_\mathrm{R}$); moreover because the soliton spectrum bridges the interval from its central frequency to the pump frequency, the parameter $\Omega_\mathrm{R}$ is a key influencer of the solitons' bandwidth (and hence duration). Assuming that dispersion can be truncated at third order ($\beta_k = 0$ for $k>3$), Eqs.~\eqref{PM} and~\eqref{GVM} yield (for $q = 0$)
\begin{equation}
\Omega_\mathrm{R} = -\frac{3\beta_2}{2\beta_3}. \label{OmR}
\end{equation}
We thus see that a large group-velocity dispersion coefficient $\beta_2$ at the pump frequency $\omega_\mathrm{P}$ is conducive for obtaining Raman solitons with broad bandwidths (short durations). Pragmatically, $\beta_2$ can be increased simply by shifting the pump frequency away from the zero-dispersion wavelength of the resonator. Figure~\ref{fig3}(a) shows results obtained in a 1-m-long resonator that illustrate these points. Here the solid curves show optical spectra measured for three different pump center wavelengths, hence different values of the group-velocity dispersion coefficient $\beta_2$. As predicted by Eq.~\eqref{OmR}, shifting the pump to shorter wavelengths (with larger $\beta_2$) shifts the Raman soliton center frequency away from the pump, whilst concomitantly yielding an overall broader spectrum. Also shown as dashed curves in Fig.~\ref{fig3}(a) are results from numerical simulations.

\begin{figure*}[!t]
	\centering
	\includegraphics[width = \textwidth, clip=true]{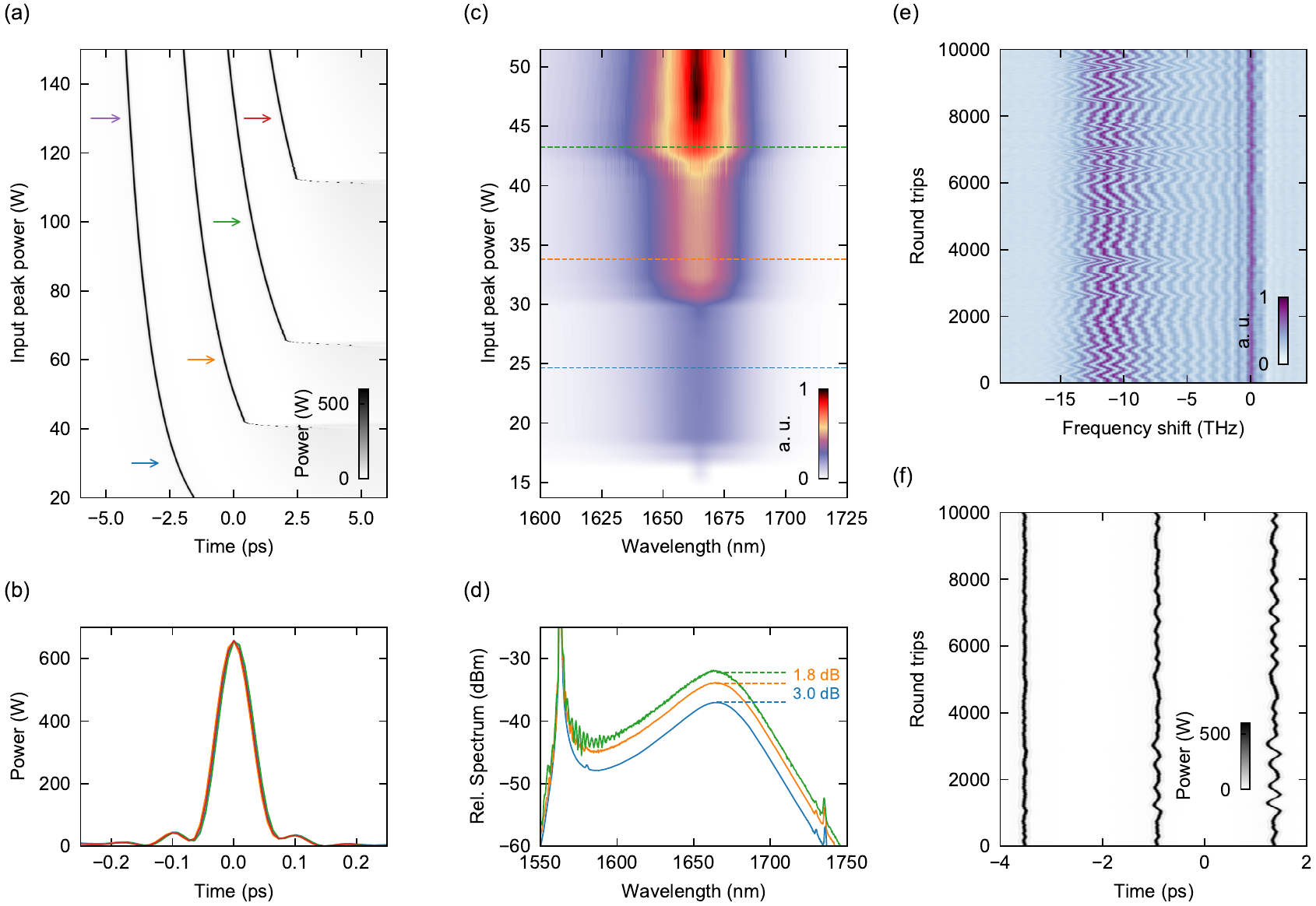}
	\caption{\textbf{Multi-soliton operation.} (a) Numerical simulation results, showing the sequential emergence of new Raman solitons in a 1-m-long resonator as the input peak power is continuously increased [for other parameters, see Methods]. (b) Overlaid temporal intensity profiles of five Raman solitons indicated by the colored arrows in (a). All of the profiles overlap, highlighting that all of the solitons exhibit identical characteristics. (c) Experimentally measured optical spectra for various input peak powers in a 1-m-long resonator. Line plots corresponding to the blue, orange, and green dashed curves are shown in (d); these highlight the 3~dB and 1.8~dB increase in the soliton spectral amplitude when respectively transitioning from one to two, and from two to three solitons. Note: spectra in (c) are plotted in linear scale and focus on the Raman soliton for clarity, whilst the spectra in (d) are plotted in log scale and include the pump frequency. (e) Results from dispersive Fourier transform measurements made for a 0.5-m-long resonator that show the optical spectrum for a two-soliton state each round trip. Jitter in experimental parameters causes the spectral interference to jitter, which results in averaged spectra with no visible fringes [see (d)]. (f) Numerical simulation of a three-soliton state [parameters as in (a) with pump power 80~W], showing how timing jitter with 5~fs standard deviation in the pump pulse repetition rate gives rise to jitter in the relative separation of the Raman solitons, with the trailing solitons experiencing larger jitter.}
	\label{fig4}
\end{figure*}

Theoretical analyses [see Methods] reveal that, in addition to $\Omega_\mathrm{R}$, the soliton's overall bandwidth also depends upon the resonator's characteristic dispersive timescale $\tau_\mathrm{s} = \sqrt{|\beta_2|L\mathcal{F}/(2\pi)}$, where $\mathcal{F}$ is the finesse of the resonator; smaller $\tau_\mathrm{s}$ generally yield broader soliton spectra. This point is demonstrated in Fig.~\ref{fig3}(b), in which we show measurements performed under conditions where the resonator length is kept constant but finesse is varied [see Methods].

The scaling laws discussed above suggest that different pump-resonator configurations for which the parameters $\Omega_\mathrm{R}$ and $\tau_\mathrm{s}$ are constant can support Raman solitons with identical characteristics. In particular, for a given chromatic dispersion, solitons with identical bandwidth (hence duration) can be sustained across systems for which the product between the round trip length $L$ and finesse $\mathcal{F}$ is an invariant. This scaling is illustrated in Fig.~\ref{fig3}(c). Here we show experimentally measured and numerically simulated optical spectra at the output of resonators with three different round trip lengths: 0.50~m, 1.02~m, and 2.65~m. For each case, we fine-tune the cavity finesse so as to obtain an approximately constant product $L\mathcal{F} \approx 250~\mathrm{m}$.  As can be seen, the optical spectra measured display high similarity, thus corroborating the scaling law. On the other hand, the corresponding RF spectra [Fig.~\ref{fig3}(d)--(f)] make it clear that the Raman solitons are generated at very different repetition rates due to the different lengths of the resonators considered. It is worth emphasising that, based on numerical modelling, the Raman solitons of Fig.~\ref{fig3}(c) and (d) have a duration of about 55~fs: the results accordingly highlight how the scheme can be used to generate ultrashort pulses across resonator length scales (and hence repetition rates).

In obtaining the results shown in Figs.~\ref{fig3}(b) and~(c), the peak power of the injected pulses was increased with reducing finesse so as to maintain net Raman gain at the group-velocity-matched component. Extensive simulations show that, whilst the Raman soliton existence depends on the pump power, their characteristics do not: provided that the input power is sufficiently large to provide net Raman gain, the same soliton characteristics ensue. However, as illustrated by the numerical simulation results shown in Figs.~\ref{fig4}(a) and (b), as the pump power is increased, the Raman gain available may exceed the amount that a single soliton can deplete [see Fig.~\ref{fig2}(d)], resulting in the sequential appearance of new (identical) solitons trailing the first one. The Raman soliton characteristics also appear to be insensitive to the detuning between the pump carrier frequency and a cavity resonance, which is in stark contrast to conventional dissipative Kerr cavity solitons~\cite{coen_universal_2013}. In fact, for the Raman solitons, experiments and simulations suggest the detuning plays a comparatively similar role as the input pump power: both influence the intracavity field at the pump frequency prior to Raman soliton formation, which in turn dictates the net gain available for soliton (or multi-soliton) formation.

Multi-soliton operation can be readily observed also in our experiments. The pseudocolor plot in Fig.~\ref{fig4}(c) shows the measured evolution of the Raman soliton spectrum as a function of injected pump power, whilst Fig.~\ref{fig4}(d) shows selected line spectra in more detail.  As can be seen, the spectral density around the soliton feature sequentially increases in discrete, equal steps. It is interesting to note that, whilst the presence of two temporally-separated solitons would be expected to yield interference fringes across the optical spectrum, such fringes are only observed around a narrow region about the pump. To gain more insights, we used the dispersive Fourier transform technique [see refs.~\cite{tong_fibre_1997, goda_dispersive_2013, runge_observation_2015} and Methods] to measure the optical spectrum at the resonator output from round trip to round trip. The results, shown in Fig.~\ref{fig4}(e), reveal that spectral interference is present across the entire spectrum but with the fringes moving about erratically over timescales of several photon lifetimes. As the optical spectra shown in Fig.~\ref{fig4}(d) are obtained from time-averaged measurements, the interference pattern is largely washed away.

The drifting of the spectral interference pattern [Fig.~\ref{fig4}(e)] suggests that the solitons' relative phase and/or temporal separation is fluctuating in our experiments, in a manner similar to previous observations of vibrating soliton molecules in mode-locked lasers~\cite{herink_real-time_2017, krupa_real-time_2017}. Numerical simulations reveal that, in our configuration, this phenomenon originates from noise in one or more of the system parameters (e.g. timing jitter or amplitude noise of the pump pulses). Specifically, because the temporal position of a trailing soliton is dictated by regeneration of the Raman gain from the depletion induced by all of the leading solitons, any fluctuation in the background field at the pump frequency (which acts as the Raman pump) results in jitter of the solitons' relative position. The same line of thought suggests that, in a $N$ soliton complex, the first (leading) soliton will experience the smallest jitter whilst the last ($N^\mathrm{th}$) soliton will experience the largest jitter (being impacted by the gain depletion of all the leading solitons). This hypothesis is corroborated by our simulations [Fig.~\ref{fig4}(f)].

\section{Discussion}

Our work demonstrates that phase-coherent pulsed driving of ``passive'' silica Kerr resonators enables the generation of coherent and broadband optical spectra and corresponding ultrashort pulses thanks to stimulated Raman scattering. In stark contrast to conventional temporal CSs of driven Kerr resonators, this new type of dissipative Raman solitons can exhibit pulse durations well below 100 femtoseconds even when realised in macroscopic fibre resonators with modest finesse and nonlinearity. To the best of our knowledge, these are the shortest pulses ever generated in resonators made from a single piece of commercially-available optical fibre (active or passive). We have demonstrated that the ultrashort Raman solitons can be deterministically generated in resonators of varying lengths, thus providing an attractive avenue for the generation of optical frequency combs with desired line spacing. We have studied the solitons' scaling laws, identifying conditions that are conducive for increasing (reducing) the solitons' bandwidth (duration): simulations indicate that durations below 40~fs should be attainable using commercially available fibers [see Supplementary Fig.~3]. Finally, whilst our experiments have been performed in fibre-based resonators, we expect that the dissipative Raman solitons can also be generated in synchronously-pumped chip-scale silica resonators, thus enabling an interesting new avenue for the generation of ultra-broadband frequency combs with small comb line spacing in photonic integrated platforms~\cite{lee_chemically_2012, yang_broadband_2016, fujii_dispersion_2020}.

\section*{Methods}

\small

\subparagraph*{\hskip-10pt Additional experimental details and parameters.} The fibre Fabry-P\'erot resonators used in our experiments are formed around two conventional FC/PC fibre connectors with standard single-mode-fibre (SMF-28) core that are dielectrically coated to produce high reflectivity over a broad band. The resonators are formed by connecting the two coated connectors with a segment of Corning MetroCor non-zero dispersion-shifted optical fibre. Resonators of various lengths can be straightforwardly achieved by simply changing the length of the fibre segment connecting the two mirrors, and it is predominantly for this reason that we chose a Fabry-P\'erot rather than a ring architecture.  We measure the finesse of the resonators by scanning a narrow-linewidth cw laser across multiple cavity free-spectral ranges and by fitting Lorentzian resonance shapes onto each adjacent resonance pair. For each pair, we obtain a statistical sample of the finesse by dividing the resonance separation with the average resonance linewidth; the final estimate for the finesse is obtained as the average of about 1000 resonance pairs. We measured the mirror transmission (i.e., coupling) coefficient into the resonator to be about $\theta = 0.002$. We note that the coupling is not lossless: the maximum resonator finesse that we have obtained is around 500, which is considerably smaller than the finesse accounting solely for the coupling ($\mathcal{F}_\mathrm{max}=\pi/\theta\approx1600$). We suspect the losses arise mostly due to imperfect contact between the connectors of the mirrors and the MetroCor fibre, and from the mode-mismatch between SMF-28 and MetroCor. We also note that the finesse is systematically tuned in our experiments by placing a small segment of the optical fibre that forms the resonator within an inline polarization controller, which introduces micro-bends into the fibre, and thus increases the resonator losses.

The dispersion in the resonators used in our work arise from the optical fibre and from the dielectric mirrors. We estimate the overall resonator dispersion from desynchronization scan data similar to the one shown in Fig.~\ref{fig1}(b). Specifically, we extract the dispersion coefficients by fitting the group-velocity-matching condition [Eq.~\eqref{GVM}] to the experimentally observed positions of the Raman components as the desynchronization is varied. We then further constrain the dispersion coefficients by fitting numerical simulations of the Raman solitons to selected experimental measurements. This two-step procedure yields the following second- and third-order dispersion coefficients at the wavelength of 1562.5~nm: $\beta_2 = 7.15~\mathrm{ps^2km^{-1}}$ and $\beta_3 = 1.36~\mathrm{ps^3km^{-1}}$. (Higher-orders of dispersion are negligible.) These coefficients predict the resonator zero-dispersion wavelength to be 1633~nm. We estimate (based on the quoted mode-field diameter) that the MetroCor fibre has a Kerr nonlinearity coefficient $\gamma=2.6~\mathrm{W^{-1}km^{-1}}$.

The resonators are pumped with a train of pulses that are generated by passing cw laser light through an electro-optic-comb generator comprised of one amplitude modulator and two phase modulators in cascade~\cite{xu_harmonic_2020}. The resulting electro-optic frequency comb is passed through 300~metres of dispersion compensating fibre and then through a nonlinear (soliton) compression scheme consisting of a 1-km-long segment of SMF-28. The pulses generated in this manner pass through a nonlinear amplifying loop mirror to remove residual low-level backgrounds. Before the pulses are launched into the resonators, they are amplified with an erbium-doped fibre amplifier. The repetition rate of the pump pulse train (hence the pump periodicity $t_\mathrm{P}$) is defined by an electronic clock signal that is fed into the modulators that make the electro-optic comb generator; adjusting this clock frequency allows to control the desynchronization $\Delta t = t_\mathrm{R}-t_\mathrm{P}$, where $t_\mathrm{R}$ is the reciprocal of the resonator free-spectral range at the pump wavelength. To calibrate the desynchronization (i.e., to determine the point where $\Delta t = 0$), we adjust the clock frequency until the spectrum around the pump frequency is symmetric~\cite{xu_frequency_2021}. We estimate that the uncertainty in the calibration is about 2~fs. The carrier frequency of the driving pulse train is locked at a fixed detuning from a cavity resonance using the Pound-Drever-Hall (PDH) technique. Specifically, a small portion of the cw laser is phase modulated and frequency shifted, and then coupled into a cavity mode that is orthogonally polarized with respect to the main pump. (The frequency-shifter compensates for the resonator birefringence and allows to control the value of the detuning experienced by the main pump.) At the cavity output, the cw component is isolated using a polarizing beam-splitter, detected, and fed to an active controller that generates the PDH signal and actuates the frequency of the driving cw laser.

The dispersive Fourier transform measurements shown in Fig.~\ref{fig4} were obtained by first spectrally filtering the resonator output to attenuate the pump component, then passing the resulting signal through a 100-m-long segment of dispersion compensating fibre with dispersion \mbox{$\beta_2=144.2~\mathrm{ps^2km^{-1}}$} to realise the frequency-to-time mapping. The output was then recorded in real-time on a 12.5~GHz photodetector and digitized on a 12~GHz real-time oscilloscope. A single, long time trace was recorded, divided into segments that span a single round trip, and finally concatenated vertically to yield the pseudo-color plot in Fig.~\ref{fig4}(e).

\subparagraph*{\hskip-10pt Numerical model.} Our numerical simulations are based on the generalized Lugiato-Lefever equation that includes pump-cavity desynchronization, higher-order dispersion, and stimulated Raman scattering. The equation describes the evolution of the slowly-varying intracavity electric field envelope $E(t,\tau)$ according to~\cite{wang_stimulated_2018}
\begin{align}
	t_\mathrm{R}\frac{\partial E}{\partial t} =& \left[-\alpha-i\delta_0-\Delta t \frac{\partial}{\partial \tau} + i L \hat{D}\left(i\frac{\partial}{\partial\tau}\right) \right]E + \sqrt{\theta P_\mathrm{P}} S(\tau) \nonumber \\
	& +i\gamma L \left[(1-f_\mathrm{R})|E|^2 + f_\mathrm{R}h_\mathrm{R}(\tau)\ast|E|^2 \right]E. \label{LLE}
\end{align}
Here the continuous variable $t$ is a ``slow'' time that describes the evolution of the field envelope $E(t,\tau)$ at the scale of the cavity photon lifetime, whilst $\tau$ is a ``fast'' time variable that describes the field profile over a single round trip. $\alpha = \pi/\mathcal{F}$ is half the power dissipated by the cavity per round trip, $\delta_0$ is the phase detuning between the injected pump pulse train and the closest cavity resonance (stabilised at $\delta_0 = 0$ in our experiments), $L$ is the cavity length, $\Delta t$ and $\hat{D}$ are the pump-cavity desynchronization and dispersion operator as defined in the main text, and $\theta$ is the power coupling coefficient of the coupler through which the pump pulses are injected into the cavity. $P_\mathrm{P}$ is the peak power of the pulses driving the cavity, whilst $f(\tau)$ describes the functional form of the pulse envelope. It is noteworthy that the (fast time) reference frame of Eq.~\eqref{LLE} has been chosen as the reference frame that is synchronous with the pump pulse train (but not the cavity round trip time), due to which (i) the desynchronization term $\Delta t$ appears explicitly in Eq.~\eqref{LLE} and (ii) the driving field profile $S(\tau)$ does not depend on the slow time $t$. The second row of Eq.~\eqref{LLE} describes the nonlinear effects: $\gamma$ is the Kerr nonlinearity coefficient, $f_\mathrm{R}$ is the Raman fraction, and $h_\mathrm{R}(\tau)$ is the time-domain Raman response, with the convolution $h_\mathrm{R}(\tau)\ast|E|^2$ reflecting the delayed nature of the Raman nonlinearity.

All of the simulation results shown in our work were obtained by numerically integrating Eq.~\eqref{LLE} using the well-known split-step Fourier method, and (apart from one exception to be discussed below) with parameters close to accepted literature values or values measured for the corresponding experiments (quoted in the main text or in the section above). For desynchronizations, we find that multiplying the value obtained by solving Eqs.~\eqref{PM} and~\eqref{GVM} with $q = 0$ by 1.1 typically yields a value within the soliton existence regime.  The simulations assume that the pump pulses have a Gaussian profile, such that
\begin{equation}
	S(\tau) = e^{-\tau^2/\tau_0^2},
\end{equation}
with the width parameter $\tau_0$ estimated as 5~ps.

The Raman response function used in our simulations corresponds to the analytical multi-vibrational model described in ref.~\cite{hollenbeck_Multiple-vibrational-mode_2002}. Interestingly, however, we find that reaching quantitative agreement with our experiments requires us to set the Raman fraction as $f_\mathrm{R} = 0.09$, which is half of the accepted value for fused silica. This discrepancy is not limited to a single set of parameters: we find that the value $f_\mathrm{R} = 0.09$ allows to obtain good agreement with experiments performed across numerous resonators with different lengths and finesse as well as both in Fabry-P\'erot and ring geometries, with all other parameters corresponding to measured or accepted values. In contrast, using the accepted value $f_\mathrm{R}=0.18$ does not yield simulation results that are quantitatively similar to our experiments. We suspect that the use of a reduced Raman fraction effectively lumps together the individual influences of a variety of higher-order effects that are present in the experiments but are ignored in Eq.~\eqref{LLE}, such as wavelength-dependent nonlinearity (self-steepening), wavelength-dependent losses, and polarization effects. All of such higher-order effects would tend to reduce the efficiency of stimulated Raman scattering, which we hypothesise can be effectively described by simply using a reduced Raman fraction of $f_\mathrm{R} = 0.09$.

\subparagraph*{\hskip-10pt Raman gain.}  The net Raman amplitude gain per round-trip experienced by a signal with carrier frequency down-shifted by $\Omega$ from a pump pulse can be approximately written as~\cite{agrawal_nonlinear_2019}
\begin{equation}
	g \approx \frac{g_\mathrm{R}}{2}P_\mathrm{P} - \alpha.
	\label{gain}
\end{equation}
Here the coefficient $g_\mathrm{R}$ is related to the resonator and Raman response parameters via $g_\mathrm{R} = 2\gamma Lf_\mathrm{R}\text{Im}[\tilde{h}_\mathrm{R}(\Omega)]$, where $\tilde{h}_\mathrm{R}(\Omega)$ is the Fourier transform of the Raman response function $h_\mathrm{R}(\tau)$, and $P_\mathrm{P}$ is the power of the Raman pump. By setting $g = 0$, we can readily derive an expression for the threshold pump power to achieve net Raman gain:
\begin{equation}
P_\mathrm{th} = \frac{2\alpha}{g_\mathrm{R}}. \label{Pth}
\end{equation}
The thresholds shown in Fig.~\ref{fig2}(c) and (d) were computed using Eq.~\eqref{Pth} with $\Omega$ chosen as the group-velocity-matched frequency.

\subparagraph*{\hskip-10pt Matching conditions.} We present here a brief derivation of the phase-matching and group-velocity-matching conditions given by Eqs.~\eqref{PM} and~\eqref{GVM}, respectively. When evaluated at a time corresponding to the period of the input pulse train, the phase accumulated by the intracavity field at the pump frequency after one round trip is given by $\phi_\mathrm{P} = \beta(\omega_\mathrm{P})L - \omega_\mathrm{P}t_\mathrm{P}$, where $\beta(\omega)$ is the propagation constant of the resonator. Similarly, the phase accumulated by the Raman soliton, with carrier frequency $\omega$, is  $\phi_\mathrm{S} = \beta(\omega)L - \omega t_\mathrm{P} + qL$, where $q$ describes the soliton's nonlinear phase shift. Equating $\phi_\mathrm{P} = \phi_\mathrm{S}$ yields
\begin{equation}
	\beta(\omega_\mathrm{P})L - \omega_\mathrm{P}t_\mathrm{P} = \beta(\omega)L - \omega t_\mathrm{P} + qL.
\end{equation}
Expanding the propagation constant as a Taylor series about the pump frequency $\omega_\mathrm{P}$ yields
\begin{equation}
	 - \omega_\mathrm{P}t_\mathrm{P} = \beta_1(\omega-\omega_\mathrm{P})L + \hat{D}(\Omega)L - \omega t_\mathrm{P} + qL,
\end{equation}
where $\Omega = \omega-\omega_\mathrm{P}$ and $\hat{D}(\Omega)=\beta(\omega)-\beta(\omega_\mathrm{P})-\beta_1\Omega$. Rearranging, we obtain
\begin{equation}
 \hat{D}(\Omega)L + qL = (\omega- \omega_\mathrm{P})t_\mathrm{P} - \beta_1(\omega-\omega_\mathrm{P})L.
\end{equation}
Using the fact that $\beta_1 L = t_\mathrm{R}$, where $t_\mathrm{R}$ is the resonator round trip time at the pump frequency, we have
\begin{equation}
	\hat{D}(\Omega)L + qL = (\omega- \omega_\mathrm{P})\left[t_\mathrm{P}-t_\mathrm{R}\right] = -\Omega \Delta t.
\end{equation}
Rearranging and dividing by $L$ yields the phase-matching Eq.~\eqref{PM}.

For the group-velocity-matching condition, we first note that the resonator round trip time at frequency $\omega$ can be written as
\begin{equation}
	t_\mathrm{R}(\omega) = \beta_1(\omega)L,
\end{equation}
where $\beta_1(\omega) = d\beta/d\omega$. Again expanding the propagation constant as a Taylor series about the pump frequency $\omega_\mathrm{P}$ yields
\begin{equation}
	t_\mathrm{R}(\omega) = t_\mathrm{R}(\omega_\mathrm{P}) + \hat{D}_1(\Omega) L,
\end{equation}
where $\hat{D}_1(\Omega) = d\hat{D}/d\Omega$. Equating this round trip time with the pump periodicity $t_\mathrm{P}$ yields
\begin{equation}
	\hat{D}_1(\Omega)L = t_\mathrm{P}-t_\mathrm{R}(\omega_\mathrm{P}) = -\Delta t.
\end{equation}
Rearranging and dividing by $L$ yields the group-velocity-matching Eq.~\eqref{GVM}.

When truncating the dispersion at third-order ($\beta_k = 0$ for $k>3$) and setting the nonlinear phase-shift $q = 0$, it is straightforward to solve Eqs.~\eqref{PM} and~\eqref{GVM} to yield Eq.~\eqref{OmR}. In this case, one finds that the corresponding desynchronization $\Delta t= 3\beta_2^2 L/(8\beta_3)$.

\subparagraph*{\hskip-10pt Scaling laws.} The generalized Lugiato-Lefever Eq.~\eqref{LLE} can be written in a dimensionless form by introducing the following variable transformations~\cite{wang_stimulated_2018}
\begin{align}
	t\rightarrow \frac{\alpha t}{t_\mathrm{R}}, \qquad \tau\rightarrow \frac{\tau}{\tau_\mathrm{s}}, \qquad E\rightarrow E\sqrt{\frac{\gamma L}{\alpha}},
\end{align}
where the fast time normalization timescale $\tau_\mathrm{s} = \sqrt{\frac{|\beta_2|L}{2\alpha}}$. Applying these changes of variables transforms Eq.~\eqref{LLE} into the following dimensionless form:
\begin{align}
	\frac{\partial E}{\partial t} =& \left[-1-i\Delta - d_1 \frac{\partial E}{\partial \tau} + i \hat{D}_\mathrm{N}\left(i\frac{\partial}{\partial\tau}\right) \right]E + \sqrt{X} f\left(\tau\tau_\mathrm{s}\right) \nonumber \\
	& +i \left[(1-f_\mathrm{R})|E|^2 + f_\mathrm{R}\Gamma(\tau,\tau_\mathrm{s})\ast|E|^2 \right]E. \label{LLEN}
\end{align}
Here, the normalized detuning $\Delta = \delta_0/\alpha$, desynchronization $d_1 = \Delta t/(\alpha\tau_\mathrm{s})$, driving peak power $X = \gamma L \theta P_\mathrm{P}/\alpha^3$, Raman response function $\Gamma(\tau,\tau_\mathrm{s}) = \tau_\mathrm{s}h_\mathrm{\tau}(\tau\tau_\mathrm{s})$ and dispersion operator
\begin{equation}
\hat{D}_\mathrm{N} = \sum_{k\geq 2} d_k \left(i \frac{\partial}{\partial \tau}\right),
\end{equation}
with the normalized dispersion coefficients $d_k = L\beta_k/(\alpha \tau_\mathrm{s}^k k!)$.

For a given Raman response, the solutions and dynamics of Eq.~\eqref{LLEN} depend upon the following normalized variables: the detuning $\Delta$, desynchronization $d_1$, dispersion $\hat{D}_\mathrm{N}$, fast time normalization timescale $\tau_\mathrm{s}$, driving peak power $X$, and the pump pulse profile $S(\tau)$. Extensive simulations show that the impact of the detuning $\Delta$, the driving peak power $X$, and the pump pulse profile $S(\tau)$ are limited to influencing whether Raman solitons exist, but not the soliton characteristics. Rather, the soliton characteristics appear to depend upon the desynchronization $d_1$, dispersion $\hat{D}_\mathrm{N}$ and the timescale $\tau_\mathrm{s}$. On the other hand, the desynchronization $d_1$ is straightforward to adjust (and optimise) in experiments by simply adjusting the clock frequency that defines the repetition rate of the pump pulse train; this parameter can therefore be considered unimportant from the point of view of system design. Assuming that the dispersion can be truncated at third-order, we can therefore conclude that the main resonator parameters that influence the attainable Raman soliton characteristics are the normalization timescale $\tau_\mathrm{s}$ and the third-order dispersion coefficient $d_3$. On the other hand, the latter coefficient can be written as $d_3 = -1/(\Omega_\mathrm{R}\tau_\mathrm{s})$ with $\Omega_\mathrm{R}$ defined by Eq.~\eqref{OmR}, thus highlighting that the soliton characteristics are dictated by $\tau_\mathrm{s}$ and $\Omega_\mathrm{R}$.

\section*{Acknowledgements}
\noindent We acknowledge financial support from the Marsden Fund of the Royal Society of New Zealand.

\section*{Author Contributions}

\noindent Z.L. performed all of the experiments and most of the simulations with the help of Y.X. and M.E. S.S. performed numerical simulations of soliton characteristics. X.W., W.W., X.W. and Z.Y. fabricated the dielectric mirrors used in the resonators. S.C. assisted with the analysis of results. S.G.M. helped supervise the experiments and interpret the results. M.E. supervised the overall project and wrote the manuscript with Z.L.

\section*{Data availability}

\noindent The data that support the plots within this paper and other findings of this study are available from M.E. upon reasonable request.

\section*{Competing financial interests}

\noindent The authors declare no competing financial interests.

\bibliographystyle{bibstyle2nonotes}

\newpage

\renewcommand{\figurename}{\textbf{Supplementary Figure}}
\renewcommand{\thefigure}{\textbf{\arabic{figure}}}
\setcounter{figure}{0}

\begin{figure*}[!t]
 \centering
 \begin{minipage}[c][\textheight]{\textwidth}
\includegraphics[width = 0.9\textwidth, clip=true]{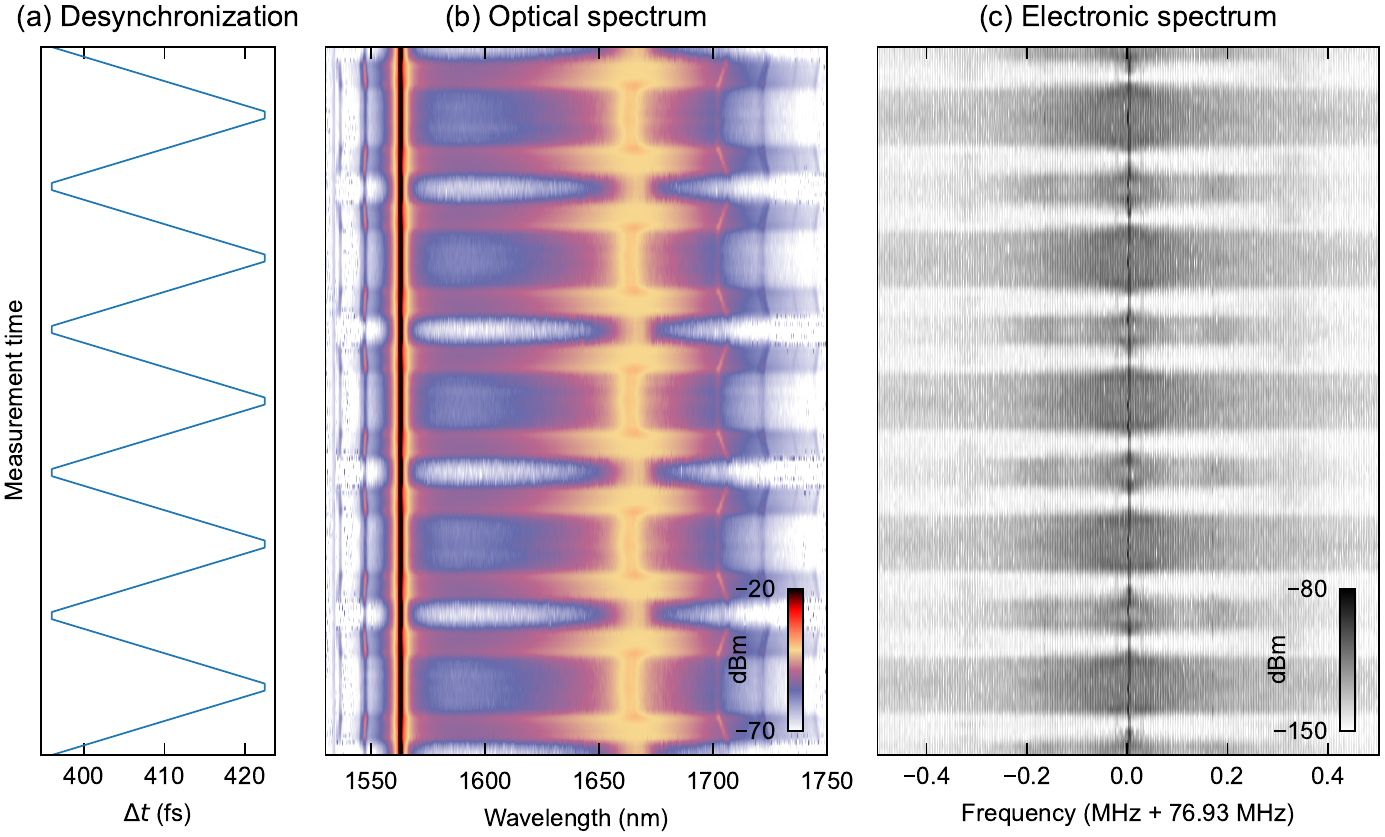}
 \caption{Experimental demonstration of deterministic Raman soliton generation. (a) The desynchronization is changed according to a triangular waveform by changing the clock frequency that defines the input pulse repetition rate. (b) and (c) show the optical and electronic spectra measured as the desyncrhonization is changed according to (a). Every time the desynchronization enters the soliton existence regime, a coherent Raman soliton emerges deterministically and spontaneously. All parameters as in Fig.~1 of the main manuscript. Note that the absolute desynchronization values needed for soliton existence differ somewhat from those observed in Fig.~1 of the main manuscript (by about 10~fs). We ascribe this to uncertainty in calibrating the zero-desynchronization point as well as to desynchronization drifts that occur during the comparatively long overall measurement times involved [e.g. 30 minutes for data shown in Fig.~1(b) and (c) of the main manuscript.].}
 \label{figS1}
 \end{minipage}
\end{figure*}

\begin{figure*}[!t]
	\centering
 \begin{minipage}[c][\textheight]{\textwidth}

	\includegraphics[width = 0.9\textwidth, clip=true]{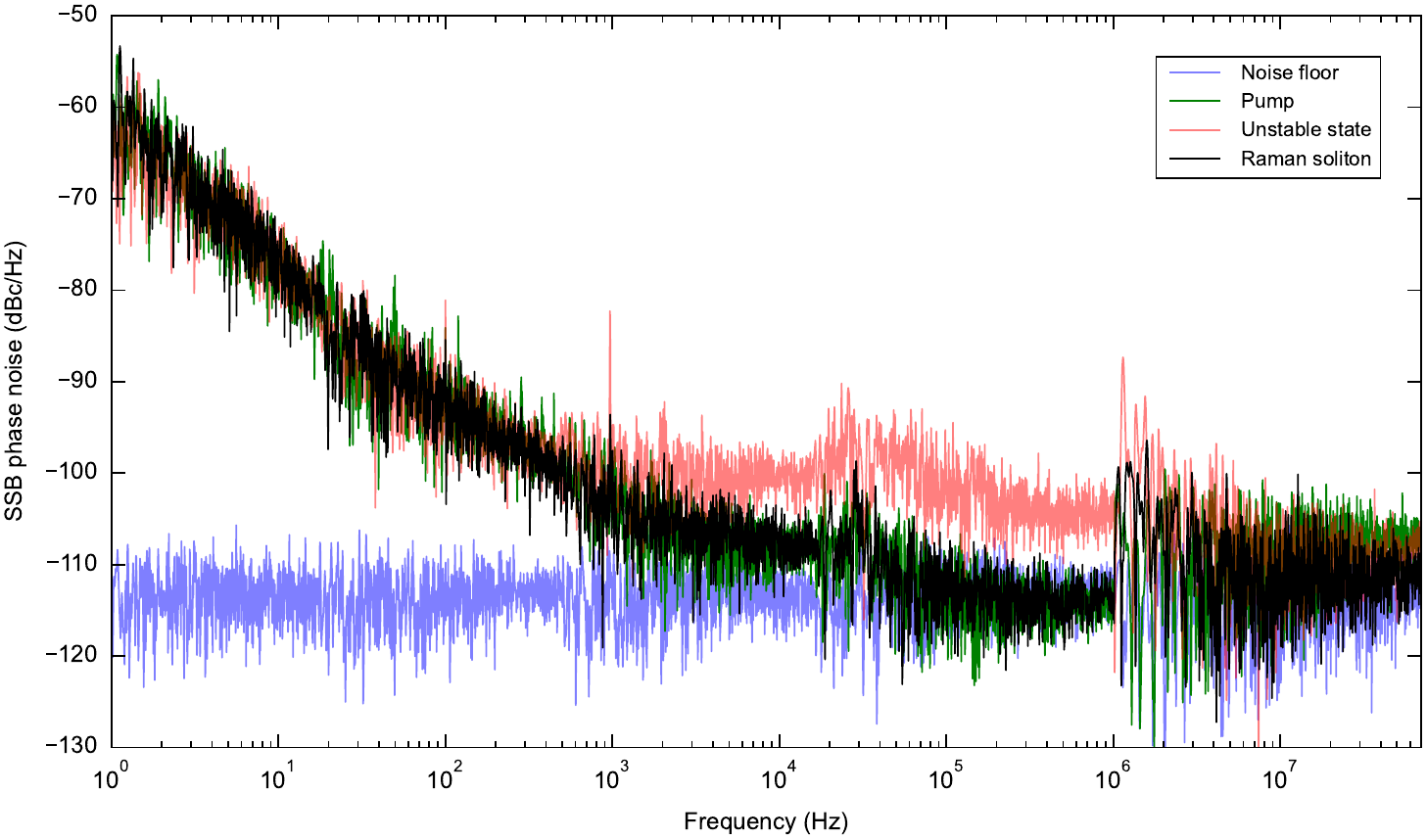}
	\caption{Single-sideband phase noise measurements for the Raman soliton state (black), the input pump (green), and an unstable state obtained by setting the desynchronization out of the Raman soliton regime (red). The purple curve shows the noise floor of the electronic spectrum analyzer. The Raman soliton phase noise is similar to the phase noise of the pump, further supporting the notion that the Raman soliton noise properties are derived from the pump.}
	\label{figS2}
\end{minipage}
\end{figure*}

\begin{figure*}[!t]
	\centering
 \begin{minipage}[c][\textheight]{\textwidth}
	\includegraphics[width = 0.9\textwidth, clip=true]{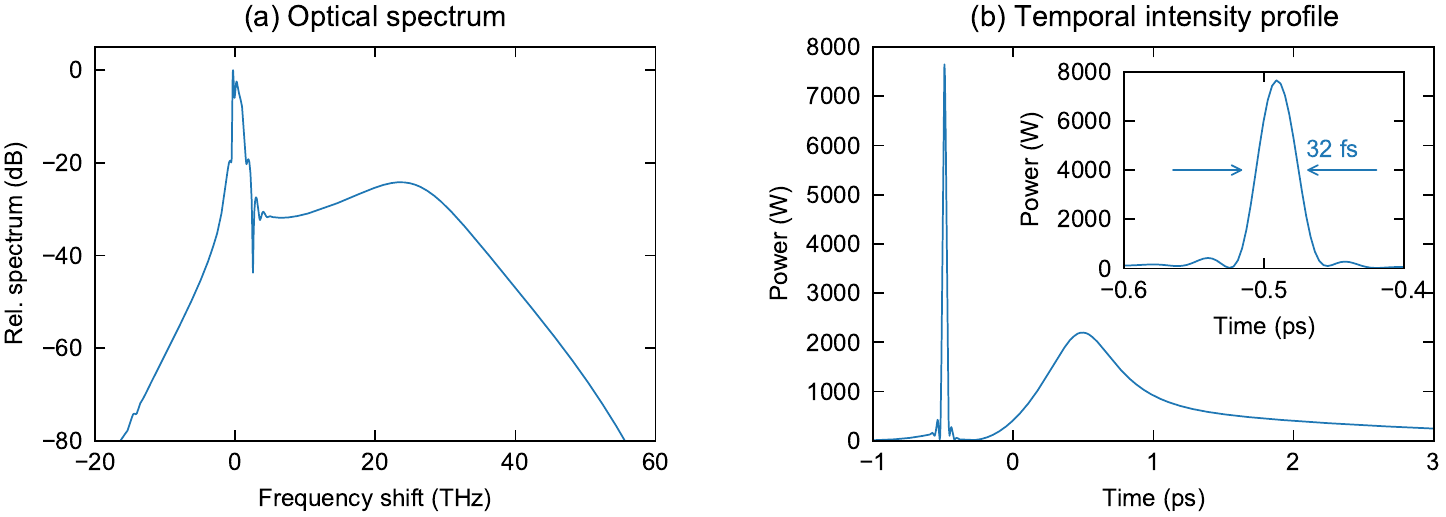}
	\caption{Numerical simulation results, showing an ultrashort dissipative Raman soliton with duration as short as 32~fs. (a) and (b) show simulated spectral (a) and temporal (b) intensity profiles of the total intracavity field. Inset in (b) shows the Raman soliton in more detail. The simulations use the generalized Lugiato-Lefever equation displayed as Eq.~5 in the main manuscript. The simulation parameters are the same as those quoted in the main manuscript expect for the following adjustments. The group-velocity dispersion coefficient was set to $\beta_2 = 14.56~\mathrm{ps^2km^{-1}}$ which corresponds to a driving wavelength of 1495~nm for the MetroCor fibre used in our work, the resonator was assumed to be 5-cm-long and to have a finesse of 400, and the Gaussian pump pulses were assumed to have an $\exp(-2)$ width of $\tau_0 = 800~\mathrm{fs}$ and a peak power of 2.5~kW. The desynchronization was set to $\Delta t = 32.15$~fs per round trip. These pump parameters (centre wavelength in particular) cannot be achieved with equipment available to us, but they are not fundamentally prohibitive.}
	\label{figS3}
\end{minipage}
\end{figure*}

\end{document}